# LLM-Based Design Pattern Detection

Christian Schindler  and Andreas Rausch
Institute for Software and Systems Engineering
Clausthal University of Technology
Clausthal-Zellerfeld, Germany
e-mail: {`christian.schindler` | `andreas.rausch`}`@tu-clausthal.de`

*Abstract*—Detecting design pattern instances in unfamiliar codebases remains a challenging yet essential task for improving software quality and maintainability. Traditional static analysis tools often struggle with the complexity, variability, and lack of explicit annotations that characterize real-world pattern implementations. In this paper, we present a novel approach leveraging Large Language Models to automatically identify design pattern instances across diverse codebases. Our method focuses on recognizing the roles classes play within the pattern instances. By providing clearer insights into software structure and intent, this research aims to support developers, improve comprehension, and streamline tasks such as refactoring, maintenance, and adherence to best practices.

*Keywords-Design Pattern detection; Large Language Model.*

## I. INTRODUCTION

Identifying design pattern instances in code is a valuable goal as it enables a deeper understanding of the structural and behavioral principles underlying software systems. By uncovering these patterns, developers and other stakeholders can gain insights into code quality, maintainability, and adherence to best practices, even in unfamiliar code bases. Automating this process can significantly reduce the time and effort required for code comprehension, facilitate knowledge transfer among teams, and improve software evolution and refactoring efforts. Furthermore, it can aid in identifying reusable components, fostering consistency, and enhancing the overall robustness of software design.

It is a challenging task due to several factors. Design patterns are often implemented with significant variations tailored to specific use cases, making consistent recognition difficult. Developers frequently deviate from canonical implementations or introduce domain-specific adaptations, complicating detection efforts. Additionally, design patterns are typically embedded implicitly within the code's structure and behavior, rather than being explicitly annotated or identifiable by a set of keywords, requiring a deep contextual understanding of the code, its dependencies, and its intent. Furthermore, applying pattern detection techniques to large, complex, and poorly documented code bases poses scalability challenges, as the sheer volume of code and intertwined dependencies can overwhelm traditional approaches and even modern tools like LLMs. These factors collectively highlight the complexity of automating design pattern identification in diverse and unfamiliar code bases.

We have defined the following Research Questions (RQs). RQ1: How can LLMs be leveraged to automatically detect and annotate design pattern instances in software codebases? RQ2: How good can LLMs detect and annotate design pattern instances in software codebases? RQ3: What are challenges/limitations faced by LLMs in identifying design pattern instances in code bases, and how can these be addressed?

This paper is organized as follows. In Section 2, we motivate the work and define the main research task. Section 3 surveys related work, while Section 4 details the experimental setup and describes the dataset. Section 5 presents the experimental results. Finally, Section 6 concludes the paper and outlines directions for future research.

## II. MOTIVATING EXAMPLE

The idea of a design pattern in software engineering is to provide a reusable, general solution to a commonly occurring problem within a given context in software design. Design patterns encapsulate best practices and proven strategies for solving these problems, offering a structured approach to building robust, maintainable, and flexible software systems. The design pattern is described on a conceptual level, defining roles and their specific responsibilities within the pattern to achieve its intended design purpose.

A design pattern instance refers to the concrete implementation of a design pattern within a specific piece of software. In this context-specific realization, the roles defined by the design pattern are embodied by individual components, such as classes, objects, or packages, which collaboratively fulfill the pattern's intended structure and behavior.

The task we want to work on is to localize such design pattern instances in code. We want to detect the design pattern applied with the respective pairs of components and their roles.

## III. RELATED WORK

### A. Design Pattern Detection

Detecting design pattern instances in existing code is a desirable task in later stages of a software system's lifecycle, particularly for maintenance purposes and to facilitate the onboarding of new developers. Various approaches have been proposed to address this challenge.

Matching-based approaches are common, such as those that rely on similarity scores [1] or graph structures [2], [3]. They have several limitations. Firstly, they often require a high degree of structural similarity between the detected instances and the target pattern. For example, they may mandate an identical number of roles or a specific implementation style. Secondly, the computational cost of the detection process can be significant, depending on the chosen matching strategy.

Furthermore, multi-stage approaches have been explored. These often involve a learning phase [4], [5] or a pattern definition phase [6] [7] prior to the actual detection of design pattern instances. Learning-based approaches in the first stage are limited by the requirement of a substantial amount of annotated training data. This data must encompass the necessary diversity of design pattern instances for the patterns of interest.

(Semi-)formal definition of design pattern structures also faces limitations. These include the expressiveness of the chosen language, the mapping of abstract concepts to language-specific constructs, and the inherent difficulty of defining patterns at this level of abstraction. The quality of this definition significantly impacts the subsequent detection phase and the accuracy of the results.

Several approaches in the literature focus on classifying source code as design pattern instances based on fixed-length inputs, such as a single class [8] [9] [10]. These methods typically work well for small or well-defined code segments where the scope is narrowly confined. However, they may fail to capture the broader context of how multiple classes or modules interact, leading to fragmented or incomplete representations of the software. In an attempt to handle larger portions of code, some methods aggregate metrics over entire modules or files—such as by summing values into a fixed-size feature vector [11]. While this strategy can reduce dimensionality and simplify processing, it tends to obscure important semantic and structural details because numerous code properties get merged into a single set of features. FeatRacer [12] addresses feature location by combining manual annotations with automated suggestions. It uses machine learning to predict missing feature recordings and guide developers, resulting in more accurate and complete feature information.

### B. Large Language Models

Large Language Models (LLMs) are advanced language models that excel at understanding and processing human language [13]. They handle tasks with minimal examples, especially involving text and data tables [14]. With larger models the performance increases [15]. Generative Pre-trained Transformers (GPTs) are powerful transformer-based models with broad applications (e.g., in education, healthcare) and challenges like high computational demands, interpretability, and ethical issues [16]. Extensively studied in computer science, GPTs relates to AI, language processing, and deep learning, as Cano [17] noted. It also aids engineering design by generating novel ideas, as Gill [18] and Zhu [19] discovered.

## IV. Experiment

### A. Data Set of Design Pattern instances

The design pattern instance data used in this work was originally collected and published by [20], and has since been adopted in various subsequent studies for benchmarking. In addition to annotating design pattern instances, the authors also provided specific versions of the corresponding source code. In this paper, we focus on instances of the Composite design pattern.

### B. Experimental setup

*1) Data Preparation:* During the data preparation phase, we thoroughly examined the available annotated design pattern instances to ensure their suitability for analysis. As part of this process, we identified and filtered out instances where not all associated source code files were accessible, as incomplete datasets could compromise the reliability of subsequent steps. Furthermore, we implemented a basic yet essential data cleaning procedure to improve the consistency of the annotations. This involved trimming unnecessary blank spaces in the XML files containing the ground truth annotations, ensuring that formatting inconsistencies did not introduce errors or ambiguities in the later stages of the workflow.

*2) Prompt Preparation:* As illustrated in the motivating example, we are providing the source code of one example instance of the design pattern with the corresponding annotation of the pattern. In addition, we are providing a second code sample, with the prompt. Listing 1 shows the prompt structure and where the described parts are located in the prompts. In a second prompt message (not listed) the source code snippet of the second code sample is provided. In addition to following the prompt structure, we further need to consider limiting factors in plugging in the instances into the prompt template. A first restriction for the prompts is the available token size of the LLM. We used ChatGPT-4 and the predecessor ChatGPT-3.5, both with a limit of 128k tokens.

Because this represents the upper bound, our interaction with the LLM must include both the prompt we provide and the expected answer, which will be much smaller than the input portion. To determine which classes to include in the provided example snippet and in the snippet for which the design pattern needs to be identified, we first identify the common root package shared by all classes participating in the ground truth annotation. For instance, the classes *a.b.c.d.ClassA* and *a.b.e.f.ClassB* share the common root package *a.b.*.

Table I provides a comprehensive overview of all remaining examples of the Composite design pattern after the Data Preparation step. For each example, it includes details about the originating project, the identified common root package, and the number of classes that (i) actively participate in the design pattern and (ii) belong to the common root package. To further optimize the input size, we have removed all comments from the classes. This reduction ensures that only essential code components remain, making the input more concise and focused for subsequent analysis and processing.

The output of this step is a collection of prompt messages. Each prompt includes a dedicated example containing a source code snippet paired with its ground truth annotation for the design pattern instance, as well as a second source code snippet from a different Java project. All permutations of the two source code snippets are considered, provided their combined token count falls within the token limit of the LLM. The order of the pairings is significant, as one snippet serves as the provided example while the other is the source code sample to be analyzed and annotated.

Listing 1. Prompt Structure of the first message

```
You are a skilled software architect. Your task
is to identify design patterns in source code
snippets and create XML annotations for them.

### Instructions:
1. Wait for a subsequent prompt that will
   contain the source code to analyze.
2. Once the source code is provided, analyze it for
   the presence of the specific code snippet _.
3. If the design pattern is found, create a valid
   XML annotation for each instance
   within the snippet.
4. If no pattern is found, simply respond with
   the text: `"No instance found."`

Important:
- Do not provide any additional explanations,
  outputs, or analyses beyond what is requested.
- If the pattern is found,
  only output the XML annotation(s).
- If the pattern is not found, only output
  `"No instance found."`

### Example:
Source Code Snippet:
File: junit.ui.TestRunner
```java
package junit.ui;

 public class TestRunner extends
junit.awtui.TestRunner {
        public static void main(String[] args) {
                new TestRunner().start(args);
        }
}```
[MORE CODE FILES]

XML Annotation:
```xml
<?xml version="1.0" ?>
<microArchitecture
    number="68" designPatternName="...">
    <roles>[...]</roles>
</microArchitecture>```
```

*3) Utilization of the LLM:* The actual prediction task is performed using ChatGPT3.5 and 4, though other LLMs could also be used due to their similar handling and functionality. Each prompt is presented to the LLM, and the generated output, which is expected to be in a valid XML format, is saved. The output is stored alongside the respective example and the design pattern expected to be found, ensuring traceability and alignment for subsequent analysis.

*4) Post processing:* We analyzed the responses provided by the LLM and systematically extracted relevant XML annotations whenever they were available as the output of the model. In cases where the LLM did not provide any annotations, it responded with a clear answer as instructed (Instruction 4 in Listing 1 and respectively treated this as no design pattern instance has been identified within the analyzed context. This approach allowed us to differentiate between cases where design patterns were explicitly recognized and cases where their absence was confirmed based on the LLM's output.

## V. RESULTS

Table II presents a confusion matrix (left hand side) illustrating the roles associated with the *Composite* design pattern.

Each column of the matrix corresponds to a distinct role, as well as to the additional categories *No Role* and *Hallucinated Class*. The rows follow a similar structure, with the key difference being the use of *Hallucinated Role* in place of *Hallucinated Class*. Within each cell, the matrix reports the total number of occurrences in which the LLM predicted a particular role (column) given a specific ground-truth role (row). For example, the cell at the intersection of the *Leaf* row and the *Composite* column indicates that, on four separate occasions, a class that was actually labeled as *Leaf* was misclassified by the LLM as *composite*.

The right hand side of the table display the total number of classifications made for each role and the corresponding Precision with and without the hallucinated predictions of the LLM. Likewise, the bottom two rows (twice) of the table show the total number of ground-truth occurrences of each role, along with the corresponding Recall. The cells at the intersection of these totals represents the overall number of entries in the confusion matrix.

Hallucination, a well-documented behavior in LLMs, involves blending prompt elements or generating information unsupported by any source. Our experiments also exhibited this phenomenon. In particular, we observed the model inventing classes not present in the code snippets—an occurrence we term hallucinated classes. Beyond that, we noted a second form of hallucination: the creation of new roles not found in any of the ground-truth examples. Both forms of hallucination are relatively straightforward to detect, as we have direct access to the complete set of valid labels and the classes embedded in the source code snippets. This comprehensive dataset enables a thorough verification process: we can systematically compare the model's predictions against the known ground-truth labels and source code elements. By doing so, it becomes evident when the model invents new classes that never appear in the provided snippets, or assigns roles not included in any of the reference examples. With this approach, any fabricated or nonexistent labels are quickly exposed, allowing us to confidently identify and document the instances in which the LLM strays beyond the established boundaries of the given code context.

In addition to the multi-class confusion matrix, we also report the results for individual prompts (as described in Section "Experimental Setup") in Table IV and Table V. In this evaluation, all roles—except for *No Role*—are considered as positive prediction classes, while *No Role* is treated as the negative class.

We define the standard confusion matrix terms as follows: True Positive (TP): The model correctly identifies that a given source code element serves a specific role in a design pattern instance. True Negative (TN): The model correctly identifies that a given source code element does not participate in any role of the design pattern instance. False Positive (FP): The model incorrectly classifies a source code element as fulfilling a particular role, even though it does not. False Negative (FN): The model incorrectly concludes that a source code element does not fulfill any role when, in fact, it does have a defined

| Instance | Project | Common Root Package | # Classes Participating in Pattern instance | # Classes in Root Package |
|---|---|---|---|---|
| 4 | QuickUML 2001 | | 17 | 217 |
| 65 | JUnit v3.7 | junit | 39 | 94 |
| 75 | JHotDraw v5.1 | CH.ifa.draw | 35 | 155 |
| 98 | MapperXML v1.9.7 | com.taursys | 29 | 234 |
| 129 | PMD v1.8 | net.sourceforge.pmd.ast | 3 | 108 |
| 143 | Software architecture design patterns in Java | src.COMPOSITE | 5 | 5 |

| | Hallucinated Class | Client | Component | Composite | Leaf | No Role | With Hallucination Classification Overall | Precision | Without Hallucination Classification Overall | Precision |
|---|---|---|---|---|---|---|---|---|---|---|
| Hallucinated Role | 0 | 0 | 1 | 0 | 0 | 3 | 4 | - | - | - |
| Client | 38 | 5 | 1 | 0 | 2 | 24 | 70 | .071 | 32 | .156 |
| Component | 0 | 0 | 8 | 1 | 0 | 1 | 10 | .800 | 10 | .800 |
| Composite | 1 | 0 | 0 | 4 | 12 | 10 | 27 | .148 | 26 | .154 |
| Leaf | 13 | 0 | 0 | 4 | 17 | 32 | 66 | .258 | 53 | .321 |
| No Role | 0 | 23 | 5 | 16 | 185 | 1064 | 1293 | .823 | 1293 | .823 |
| Truth overall | 52 | 28 | 15 | 25 | 216 | 1134 | 1470 | | | |
| Recall | - | .179 | .533 | .160 | .008 | .938 | | | | |
| Truth overall | - | 28 | 14 | 25 | 216 | 1131 | | | 1414 | |
| Recall | - | .179 | .571 | .160 | .008 | .941 | | | | |

| | No Role | Client | Component | Composite | Leaf | Classification Overall | Precision |
|---|---|---|---|---|---|---|---|
| Client | 13 | 9 | 1 | 3 | 8 | 34 | .265 |
| Component | 1 | 0 | 10 | 0 | 0 | 11 | .910 |
| Composite | 5 | 0 | 0 | 12 | 12 | 29 | .414 |
| Leaf | 34 | 2 | 0 | 1 | 45 | 82 | .549 |
| No Role | 1079 | 17 | 4 | 11 | 150 | 1261 | .856 |
| Truth overall | 1132 | 28 | 15 | 27 | 215 | 1417 | |
| Recall | .953 | .333 | .833 | .522 | .209 | | |

role according to the ground truth annotation.

By applying this approach, the Precision, Recall, and F1 metrics reported focus solely on assessing the model's ability to correctly identify the roles within the design pattern instances. All classes from the code snippets that are not part of the design pattern instance serve as the negative (prediction) class, ensuring that the evaluation focuses on the roles of the design pattern. In evaluating the performance of classification models, Precision, Recall, and F1 Score are key metrics that provide insight into the model's accuracy in distinguishing between the classes. These metrics are particularly important in imbalanced datasets where a models overall accuracy might be misleading.

Table IV shows the different prediction runs with ChatGPT 3.5 as a row each, with the following information. *Run* is a running number to refer to the individual prediction runs with *TP, TN, FP*, and *FN* predictions, alongside the bespoken metrics. On the right side we reported the same metrics after removed the hallucination introduced by the LLM. Hallucination occurred in three runs (2, 3, and 4). Removal of the hallucination lead to increased Precision, F1 Score and Accuracy in those runs. Four runs (i.e., 11, 12, 13, and 14) did not detect a design pattern annotation but instead responded with the no design pattern found response.

Table V lists the prediction runs with ChatGPT 4, the noticeable difference to Table IV is, that no hallucination occurred. In a consequence for each run their is only one set of metrics to be reported. In addition ChatGPT 4 has been better in 7 out of 14 runs for Precision and Accuracy, in 8 cases for F1 Score and in 9 cases for Recall, compared to ChatGPT 3.5.

Table VI provides additional details on each of the 14 runs and includes the corresponding design pattern instance IDs for both the example and the target, referencing the IDs in Table I. In this context, the example is the annotated instance, while the target is the instance for which only source code was provided via prompts. The subsequent columns show the number of role annotations in the ground truth (for both Example and Target) and those predicted by the LLMs ChatGPT 3.5 and 4.

We noticed that in all cases where the LLMs failed to identify a design pattern annotation, the same example (143) has been used. A notable aspect of this example is that the source code snippet includes only those classes that participate in the design pattern (see Table I), with no additional classes. It appears that providing the snippet within a broader implementation context helps the LLM better distinguish which classes are relevant to a design pattern instance.

Examining the confusion matrix shows that Precision and Recall vary considerably across different roles. The highest scores for both Precision and Recall occur for the role *Component* (keeping the *No Role* label out of scope), likely because each design pattern instance has exactly one class labeled with this role, and every successful prediction also identified exactly one such class. By contrast, performance for the other roles is poorer. As shown in Table VI, the number of classes labeled with many examples differs widely among examples. In the case of *Client*, for instance, some instances lack this role entirely.

TABLE IV. Results for each prediction run with ChatGPT 3.5

| Run | With Hallucination | | | | | | | | Without Hallucination | | | | | | | |
|---|---|---|---|---|---|---|---|---|---|---|---|---|---|---|---|---|
| | TP | FP | FN | TN | Precision (P) | Recall (R) | F1 | Accuracy (A) | TP | FP | FN | TN | P | R | F1 | A |
| 1 | 4 | 5 | 36 | 53 | .444 | .100 | .163 | .582 | 4 | 5 | 36 | 53 | .444 | .100 | .163 | .582 |
| 2 | 12 | 17 | 23 | 117 | .414 | .343 | .375 | .763 | 12 | **6** | 23 | 117 | **.667** | .343 | **.453** | **.816** |
| 3 | 3 | 77 | 37 | 39 | .037 | .075 | .050 | .269 | 3 | **33** | 37 | 39 | **.083** | .075 | **.079** | **.375** |
| 4 | 2 | 2 | 38 | 54 | .500 | .050 | .091 | .583 | 2 | **1** | 38 | 54 | **.667** | .050 | **.093** | **.589** |
| 5 | 1 | 29 | 16 | 171 | .033 | .059 | .043 | .793 | 1 | 29 | 16 | 171 | .033 | .059 | .043 | .793 |
| 6 | 1 | 3 | 4 | 0 | .250 | .200 | .222 | .125 | 1 | 3 | 4 | 0 | .250 | .200 | .222 | .125 |
| 7 | 3 | 4 | 2 | 0 | .429 | .600 | .500 | .333 | 3 | 4 | 2 | 0 | .429 | .600 | .500 | .333 |
| 8 | 3 | 1 | 2 | 0 | .750 | .600 | .667 | .500 | 3 | 1 | 2 | 0 | .750 | .600 | .667 | .500 |
| 9 | 3 | 1 | 2 | 0 | .750 | .600 | .667 | .500 | 3 | 1 | 2 | 0 | .750 | .600 | .667 | .500 |
| 10 | 2 | 3 | 3 | 0 | .400 | .400 | .400 | .250 | 2 | 3 | 3 | 0 | .400 | .400 | .400 | .250 |
| 11 | 0 | 0 | 17 | 200 | 0 | 0 | 0 | .922 | 0 | 0 | 17 | 200 | 0 | 0 | 0 | .922 |
| 12 | 0 | 0 | 3 | 105 | 0 | 0 | 0 | .972 | 0 | 0 | 3 | 105 | 0 | 0 | 0 | .972 |
| 13 | 0 | 0 | 29 | 205 | 0 | 0 | 0 | .876 | 0 | 0 | 29 | 205 | 0 | 0 | 0 | .876 |
| 14 | 0 | 0 | 35 | 120 | 0 | 0 | 0 | .774 | 0 | 0 | 35 | 120 | 0 | 0 | 0 | .774 |

TABLE V. Results for each prediction run with ChatGPT 4

| Run | TP | FP | FN | TN | Precision | Recall | F1 | Accuracy |
|---|---|---|---|---|---|---|---|---|
| 1 | 6 | 3 | 34 | 54 | .667 | .150 | .245 | .619 |
| 2 | 21 | 32 | 14 | 98 | .396 | .600 | .477 | .721 |
| 3 | 8 | 4 | 32 | 53 | .667 | .200 | .308 | .629 |
| 4 | 0 | 0 | 39 | 55 | 0 | 0 | 0 | .585 |
| 5 | 0 | 0 | 17 | 200 | 0 | 0 | 0 | .922 |
| 6 | 2 | 1 | 3 | 0 | .667 | .400 | .500 | .333 |
| 7 | 3 | 1 | 2 | 0 | .750 | .600 | .667 | .500 |
| 8 | 3 | 1 | 2 | 0 | .750 | .600 | .667 | .500 |
| 9 | 3 | 1 | 2 | 0 | .750 | .600 | .667 | .500 |
| 10 | 2 | 1 | 3 | 0 | .667 | .400 | .500 | .333 |
| 11 | 11 | 0 | 6 | 200 | 1.000 | .647 | .786 | .972 |
| 12 | 0 | 0 | 3 | 105 | 0 | 0 | 0 | .972 |
| 13 | 0 | 11 | 29 | 203 | 0 | 0 | 0 | .835 |
| 14 | 17 | 15 | 18 | 111 | .531 | .486 | .507 | .795 |

To address RQ1, we designed a sequence of two prompts tailored for the task. The first prompt establishes a comprehensive context by including key components: a background setting for the LLMs, a description of their assumed skill set, and explicit instructions outlining the task along with critical information to consider during problem-solving. Additionally, this prompt provides a worked example that includes an annotated design pattern instance in the desired output format, paired with its corresponding code snippet. The second prompt introduces the target code snippet for which the LLM is expected to identify and output the relevant design pattern instance. This sequential approach ensures that the LLM is both primed with contextual understanding and guided by an illustrative example before tackling the target task.

To address RQ2, we analyzed the experimental results, which revealed varying levels of quality depending on the roles within the design pattern. For instance.Furthermore ChaGPT 4 performed better compared to ChatGPT 3.5, by finding more annotations and having more correct annotations. Additionally, we observed substantial variability in quality across different runs, influenced by the pairing of the example code with unseen target code and the expected annotations.

In RQ3, we investigated the challenges and limitations of using an LLM-based approach for detecting design pattern instances. Our findings demonstrate that LLMs, such as Chat-GPT 3.5 and ChatGPT 4 in our case, are capable of processing prompts containing code snippets spanning hundreds of Java classes and accurately retrieving classes along with their correct roles. However, this approach faces several limitations.

One significant constraint is the token limit of current LLMs, which restricts scalability. In our experiments, we could only include a single example of the design pattern for a given prediction in such detail, as multiple examples would exceed the context window. Additionally, some pairings of examples and target code snippets had to be excluded because their combined token count exceeded the allowable limit.

Another challenge lies in the heterogeneity of the design pattern instances, as they vary in the number of roles and the specific roles annotated in the examples, adding complexity to the task and potentially affecting the model's performance.

If the role expected to be predicted by the LLM is not present in the provided example, it becomes an unseen role, making it challenging for the models to annotate accurately.

To address the maxing out of the token size limitation of current LLMs, an effective strategy could involve reducing the size of the code snippets provided in each prompt. By pruning the code snippets—such as focusing more on the relevant sections or removing parts that are less critical for identifying the design pattern—it may become feasible to include multiple examples of the design pattern within the context window. Presenting multiple examples could help the model better generalize across varying instances of the design pattern and potentially improve its prediction accuracy.

For instance, pruning could involve removing sections of code unrelated to the design pattern being analyzed or adjusting the balance between classes directly involved in the design pattern instances and those present in the source project but

| Run | Design Pattern Instance ID | | Composite | | | | Component | | | | Leaf | | | | Client | | | |
|---|---|---|---|---|---|---|---|---|---|---|---|---|---|---|---|---|---|---|
| | Example (E) | Target (T) | E | T | GPT4 | GPT3.5 | E | T | 4 | 3.5 | E | T | 4 | 3.5 | E | T | 4 | 3.5 |
| 1 | 4 | 65 | 1 | 2 | 1 | 1 | 1 | 1 | 1 | 1 | 14 | 37 | 6 | 6 | 1 | 0 | 1 | 1 |
| 2 | 65 | 75 | 2 | 5 | 4 | 2 | 1 | 1 | 1 | 1 | 37 | 21 | 29 | 22 | 0 | 8 | 19 | 0 |
| 3 | 75 | 65 | 5 | 2 | 1 | 10 | 1 | 1 | 1 | 1 | 21 | 37 | 7 | 17 | 8 | 0 | 3 | 52 |
| 4 | 143 | 129 | 1 | 1 | - | 1 | 1 | 1 | - | 1 | 2 | 1 | - | 1 | 1 | 0 | - | 1 |
| 5 | 143 | 98 | 1 | 1 | - | 5 | 1 | 1 | - | 1 | 2 | 22 | - | 15 | 1 | 5 | - | 9 |
| 6 | 129 | 143 | 1 | 1 | 1 | 2 | 1 | 1 | 1 | 1 | 1 | 2 | 1 | 1 | 0 | 1 | 0 | 0 |
| 7 | 98 | 143 | 1 | 1 | 1 | 1 | 1 | 1 | 1 | 1 | 22 | 2 | 1 | 1 | 5 | 1 | 1 | 4 |
| 8 | 65 | 143 | 2 | 1 | 1 | 1 | 1 | 1 | 1 | 1 | 37 | 2 | 1 | 1 | 0 | 1 | 1 | 1 |
| 9 | 4 | 143 | 1 | 1 | 1 | 1 | 1 | 1 | 1 | 1 | 14 | 2 | 1 | 1 | 1 | 1 | 1 | 1 |
| 10 | 75 | 143 | 5 | 1 | 1 | 2 | 1 | 1 | 1 | 1 | 21 | 2 | 1 | 1 | 8 | 1 | 0 | 1 |
| 11 | 143 | 4 | 1 | 1 | 1 | - | 1 | 1 | 1 | - | 2 | 14 | 8 | - | 1 | 1 | 1 | - |
| 12 | 143 | 129 | 1 | 1 | - | - | 1 | 1 | - | - | 2 | 14 | - | - | 1 | 0 | - | - |
| 13 | 143 | 98 | 1 | 1 | 5 | - | 1 | 1 | 1 | - | 2 | 22 | 2 | - | 1 | 5 | 3 | - |
| 14 | 143 | 75 | 1 | 5 | 5 | - | 1 | 1 | 1 | - | 2 | 21 | 25 | - | 1 | 8 | 1 | - |

not participating in the design pattern instance. This approach would allow for a richer variety of training examples without exceeding the token limit, thereby balancing the trade-off between example diversity and code snippet size.

## VI. CONCLUSION AND FUTURE WORK

We have demonstrated an approach for identifying design pattern instances in large codebases comprising over 200 classes presented simultaneously, using only a single example of the design pattern. This approach operates directly on the source code without requiring modifications like abstraction or feature extraction. Our results show that the model can partially retrieve design patterns with varying levels of completeness and accuracy. However, as discussed in the results section of the paper, the approach is not perfect, as it handles different roles with varying degrees of success. Moreover, the experiment we conducted was quite limited: it tested only one design pattern with two LLMs. Consequently, results for other design patterns may differ from those reported here, particularly because certain patterns could be either easier or more difficult to detect in larger codebases. In addition, the experiment was conducted as a one-shot test, providing only a single example. It would therefore be valuable to investigate how providing multiple examples might influence the model's predictive performance.